\documentclass{elsart}

\usepackage{natbib}

\usepackage{amssymb}

\begin{document}

\begin{frontmatter}



\title{Prospects for Non-Standard Interactions at MiniBooNE}


\author{Loretta M. Johnson}

\ead{ljohnson@kzoo.edu}

\author{R. Seton Williams, Laura F. Spencer, Burton J. DeWilde}

\address{Kalamazoo College, Department of Physics, Kalamazoo,
MI 49006}

\begin{abstract}
MiniBooNE is expected to soon report analysis of their closed data set, and we here consider what effect direct non-standard neutrino interactions would have on their observations.  Current direct limits on non-standard interactions make interpretation of a $\nu_e$ signal at MiniBooNE as three-flavor oscillations plus direct interactions untenable.  However, $\nu_\tau$ from non-standard interactions at the source could contribute to a $\nu_\mu$ deficit which may be observable if non-standard interactions are sizable.
\end{abstract}


\end{frontmatter}

\section{Introduction}
\label{intro}
Now that there is good evidence from reactor \citep{kamland} and accelerator experiments \citep{k2k} supporting the solar matter oscillation \citep{solar,solsk} and atmospheric oscillation solutions, \citep{atmo} the LSND \citep{LSND,LSanti} puzzle is the remaining oscillation mystery,  and soon MiniBooNE will release new evidence which should resolve that issue.  The solar and atmospheric data are consistent with a minimal three-family oscillation, with each described by a mass-squared difference and a mixing angle.  The parameter space of the remaining mixing angle has been explored by experiments, including reactor neutrino experiments, and this mixing angle is constrained to be small.  Evidence from solar, atmospheric, and reactor experiments is so compelling, that it is now common to refer to the "standard model" which includes at least two non-zero neutrino masses.  LSND does not fit within this three-angle, two mass-squared difference picture, and many authors \citep{newlsnd} have proposed alternatives.  Here we consider an alternative interpretation of the data if MiniBooNE's observations are not consistent with the standard model, and if non-standard neutrino interactions are of significant size.

Neutrinos for MiniBooNE are created when 8 GeV protons collide with a beryllium target producing 3 GeV secondary mesons which decay into mostly muon neutrinos with 0.6 $\%$ electron neutrinos expected.  The neutrino energy is around 0.5 GeV, the neutrinos travel 541 m to the detector, and they're detected primarily through quasi-elastic scattering, $\nu C \to l^- N$.  There are significant backgrounds, and the MiniBooNE collaboration has studied related experiments and planned carefully for overcoming these, including significant improvement in understanding of intrinsic $\nu_e$ from muon and kaon decays, neutral pions with only one photon ring observed, the small $\Delta \to N \gamma$ rate, and how to distinguish $\nu_e$ from $\nu_\mu$.\citep{MB}

Our question is whether non-standard neutrino interactions could be responsible if more than the expected $\nu_e$ are observed, and what would happen if non-standard interactions resulted in $\nu_\tau$.  Direct limits on wrong-flavor neutrinos in pion decay are not as strong as one might expect, and limits through $SU(2)$ symmetry from all-charged-lepton modes may be avoided in some models.   The direct limit on $\pi \to \mu \nu_{e}$ is 0.008 and no direct limit on $\pi \to \mu \nu_{tau}$ exists; indirect limits through $SU(2)$ symmetry are much stronger.\citep{pdg}  Several authors have previously studied non-standard interactions in various neutrino contexts \citep{nsi} since lepton-flavor violating interactions are present in many models that have neutrino mass and mixing.  Our model includes neutrino oscillations and a low-energy effective four-Fermi interaction that may violate lepton flavor number in a unified way. \citep{JM,JoMc}  These kinds of flavor-violating interactions may occur in a number of models, and here we remain as model-independent as possible.  In this paper we report on a three-flavor analysis of what MiniBooNE may expect to observe if the conventional solar and atmospheric solutions are correct and there are flavor-changing direct interactions.

\section{Oscillations Plus Non-Standard Interactions Model}
\label{model}

We have previously discussed a low-energy, effective four-Fermi model for including non-standard neutrino interactions with neutrino oscillations and studying their consequences.  Here we simplify the notation used in our previous papers by omitting the subscript corresponding to the flavor of the charged lepton; we assume that a meson has decayed to a muon and a neutrino, and so only a (created) neutrino flavor subscript is required.
\begin{equation}
L = 2 \sqrt{2} G_F \tan \psi_j e^{2i\phi_j} (\bar \mu \Gamma U_{j \alpha} \nu_\alpha ) [\bar d \Gamma ' u]^\dag + h.c.
\end{equation}
In this paper we consider only standard helicity and Lorentz structure (encapsulated in $\Gamma$ and $\Gamma '$) and the semi-leptonic interactions relevant to MiniBooNE.  The subscripts $j$ and $\alpha$ are allowed to be $e, \mu , \tau$ and $1, 2, 3$, respectively, for the neutrino flavor and mass eigenstates.    We use the conventional three-flavor mixing matrix, $U_{j \alpha}$, as a unitary transformation between these flavor and mass eigenstates.

Including neutrino oscillations and these non-standard interactions at the source, the flavor-change probability is:
\begin{equation}
P = | \sum_{j,c} \tan \psi_j e^{2i\phi_j} U_{jc} e^{iE_c t} U_{1c}^* |^2
\end{equation}
Here $j$ and $c$ are summed over all three flavors with $ \tan \psi_\mu = 1$ and $\phi_\mu = 0$ for the flavor expected in the standard model.  We parameterize the new interaction strengths with a magnitude given by a pseudo-angle, $\psi$, and a CP phase, $\phi$, and there is a pair of these for each of the possible unexpected neutrino flavors, $e$ and $\tau$.  Notice, then, that the detected wrong-flavor neutrino may have had several different origins:  1) the correct flavor ($\nu_\mu$) was produced but the neutrino oscillated to $\nu_e$, 2) the wrong flavor ($\nu_e$) was produced in the pion decay, and 3) the wrong flavor ($\nu_\tau$) was produced in the pion decay and oscillated to the detected wrong flavor ($\nu_e$).

\section{Non-Standard Interactions at MiniBooNE}
\label{minib}

MiniBooNE employs a large number of PMTs looking into a spherical signal region and a small number of PMTs looking outward into a veto region in order to identify Cerenkov cones created after an electron or muon neutrino interacts in the detector, creating a charged $e$ or $\mu$.  The collaboration has studied how to distinguish between them (and discriminate from background) using multiple algorithm particle identification.  The experiment is a $\nu_e$ appearance experiment with high sensitivity, but if the simplest three-neutrino mixing model is correct, MiniBooNE will not detect $\nu_e$ above background.  The experiment was devised in order to collect enough data to persuasively confirm or refute LSND's intriguing observations roughly a decade ago.

For the bulk of our study, we considered the case of $\psi_e = 0.008$ and $\psi_\tau = 0.1$.  The effects of all the CP phases (including delta from the standard model) were negligible, at most a fraction of a percent.  We used values of $\theta_{ij}$ and $\Delta m_{ij}^{2}$ consistent with solar, atmospheric, reactor, and accelerator (except LSND) data. \citep{pdg}  As shown in Figure~\ref{fig:figure1}, direct interactions are unlikely to produce enough $\nu_e$ to be observable by MiniBooNE; a flavor-change probability on the order of $10^{-5}$ would correspond to a handful of $\nu_e$ among the couple thousand expected.  Even if all of this handful were observed, they would be among the hundreds of events either containing or misidentified as $\nu_e$, a signal-to-background of less than 0.01.   This also means MiniBooNE is unlikely to be able to improve the limits on $\psi_e$.

That leaves the possibility that $\nu_\tau$ arrive in the detector.  Figure~\ref{fig:figure1} also shows that the constraints on production of tau neutrinos at the source are much weaker.  Obviously $\tau$ can't be produced through quasi-elastic scattering of neutrinos with such low energy; however, if $\nu_\tau$ are present in the beam, then the flux of $\nu_\mu$ observed would be lower by a corresponding amount.  Given that the MiniBooNE collaboration will know the number of protons on target to about 1 percent, they may be able to use muon disappearance to observe non-standard interactions or set a direct limit on $\psi_\tau$.  If MiniBooNE can perform the $\nu_\mu$ disappearance experiment, then Figure~\ref{fig:figure2} can be used to approximate either $\psi_\tau$ or their new limit on it.

\section{Discussion}
\label{disc}

We have completed a three-flavor oscillation plus non-standard interaction analysis of MiniBooNE.  We showed that while the types of new interactions we studied here are unlikely to result in an observable $\nu_e$ appearance signal, they might lead to a measurable deficit of $\nu_\mu$.

\section{Acknowledgements}
\label{ack}
We thank Doug McKay and Mary Hall Reno for discussions.  This work was supported in part by the Kamerling Fund.  We mourn the death of our collaborator, Williams.

Figure~\ref{fig:figure1} displays the low probability of a $\nu_e$ being produced with these non-standard interactions in addition to oscillations using parameters consistent with solar, atmospheric, and reactor data.  Also shown is the much higher flavor-change probability for $\nu_\tau$; however, these will be challenging to detect since $\tau$ can't be produced from such a low-energy beam.  Both flavor-change probabilities are relatively independent of $L/E$.

Figure~\ref{fig:figure2} displays how rapidly the flavor-change probability climbs as the direct non-standard interaction strength, $\psi_\tau$, increases.  Since the $\nu_\tau$ probability is approximately flat over the energy range for MiniBooNE in Figure~\ref{fig:figure1}, this second figure can be used to see what values of $\psi_\tau$ the experiment is sensitive to; if one percent too few $\nu_\mu$ are observed and the number of $\nu_e$ corresponds only to the expected background, then $\psi_\tau$ is around 0.1.




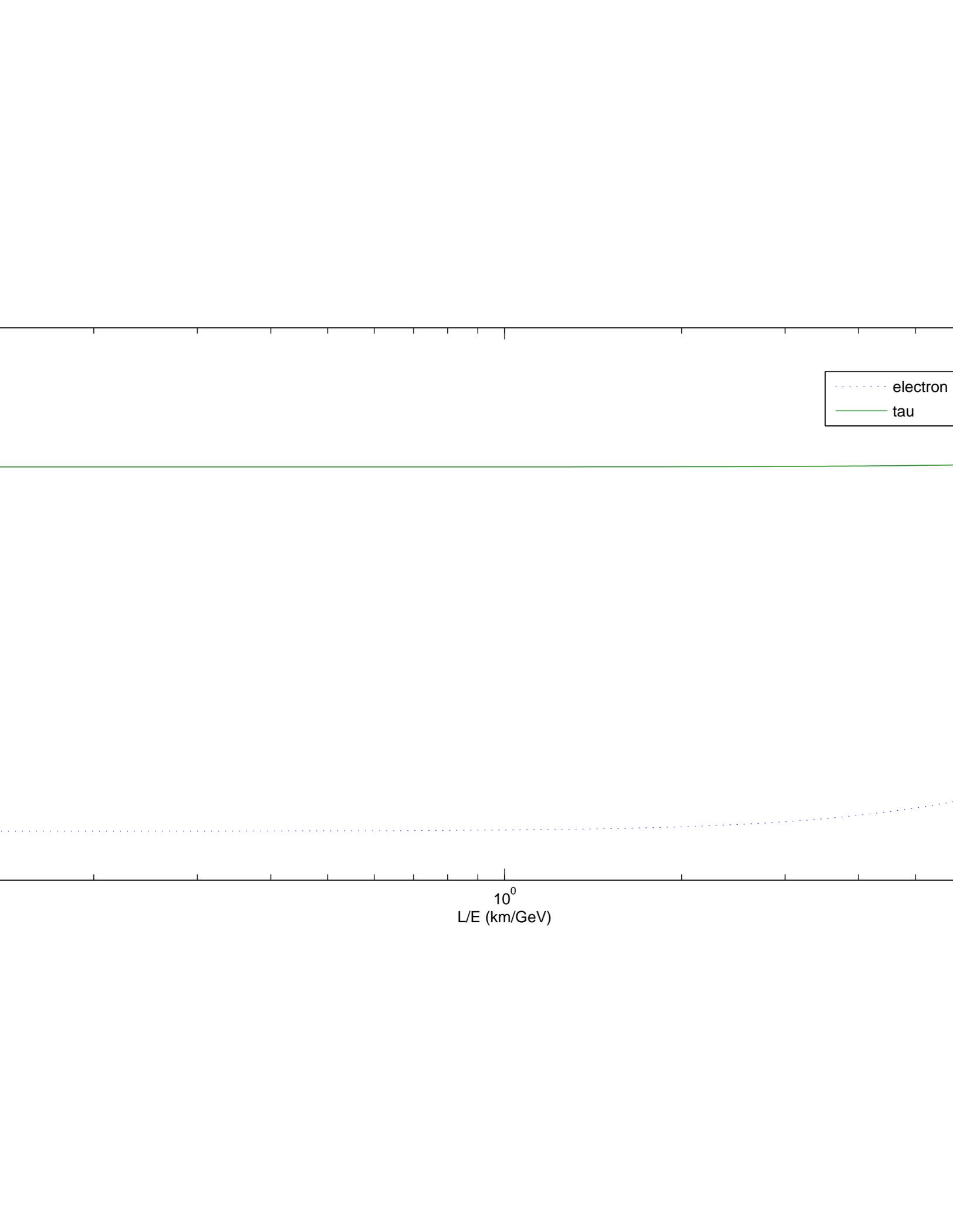
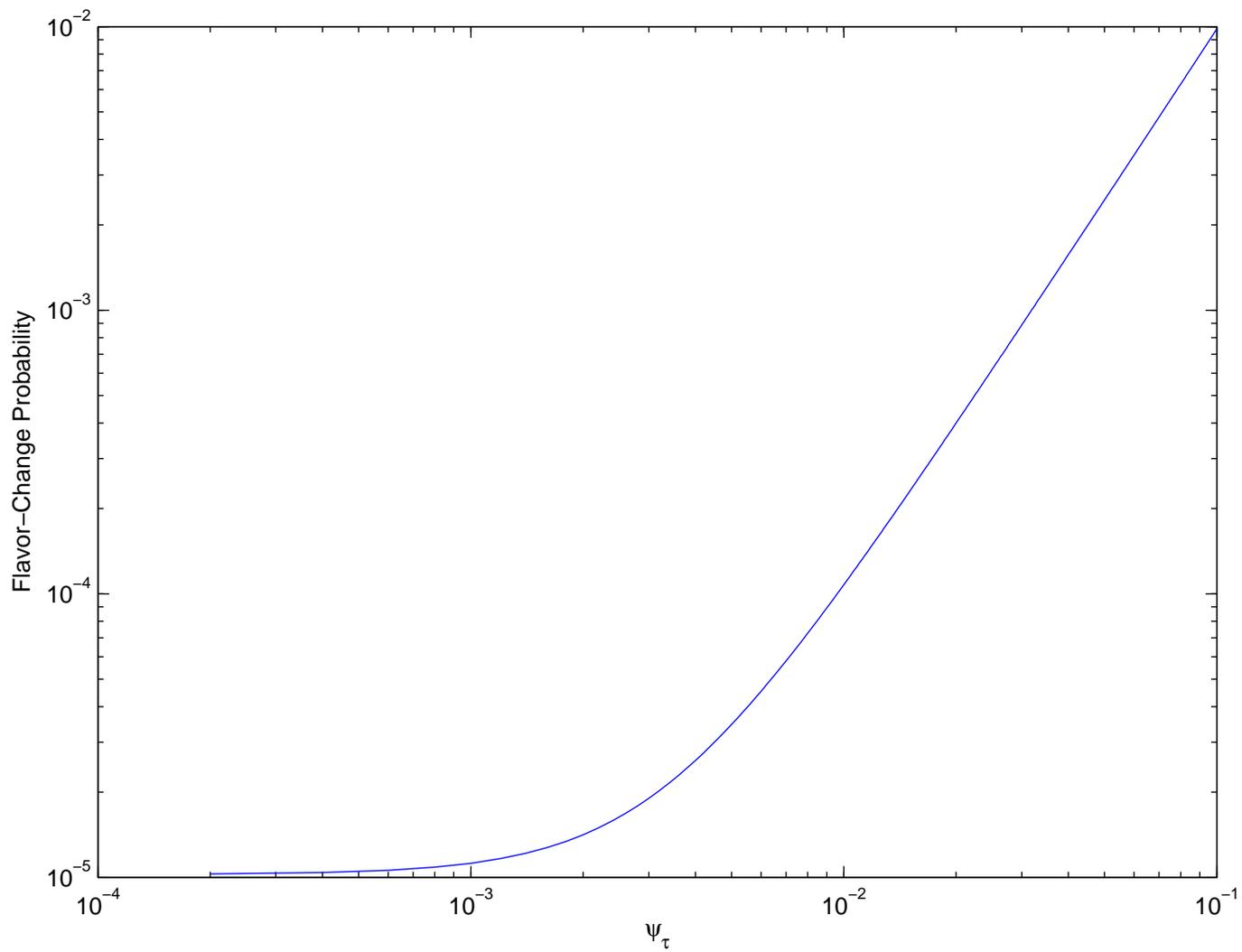
\end{document}